\documentclass[aps,pra,twocolumn,showpacs,superscriptaddress,nofootinbib,10pt,floatfix]{revtex4-2}
\usepackage{pifont}
\usepackage{bbm}
\usepackage[english]{babel}
\usepackage[utf8x]{inputenc}
\usepackage[T1]{fontenc}
\usepackage{natbib}
\usepackage{amsmath,amssymb}
\usepackage{graphicx}
\usepackage{caption}
\usepackage{subcaption} 
\usepackage{ragged2e}
\usepackage{xcolor}
\usepackage{stackrel}
\usepackage{amsfonts}
\usepackage{bm}
\usepackage[normalem]{ulem}
\usepackage{hyperref}
\usepackage{verbatim}
\usepackage{physics}
\usepackage{tabularx} 
\usepackage{array}
\usepackage[table]{xcolor}

\definecolor{rosita}{rgb}{0.97, 0.56, 0.65}
\definecolor{verde}{rgb}{0, 0.56, 0.65}
\definecolor{RedOrange}{rgb}{1, 0.5, 0}
\definecolor{byzantine}{rgb}{0.74, 0.2, 0.64}
\definecolor{brightturquoise}{rgb}{0.03, 0.91, 0.87}
\definecolor{brightube}{rgb}{0.82, 0.62, 0.91}
\definecolor{caribbeangreen}{rgb}{0.0, 0.8, 0.6}
\definecolor{amethyst}{rgb}{0.6, 0.4, 0.8}

\newcommand{\beq}{\begin{equation}}
\newcommand{\eeq}{\end{equation}}
\newcommand{\eq}{\begin{equation}}
\newcommand{\en}{\end{equation}}

\begin{document}

\title{Geometric quantifier of the incompatibility of single-particle property attribution in indistinguishable boson systems }

\author{P. Céspedes}
\email{corresponding author: paulacespedes@unc.edu.ar} 
\affiliation{Facultad de Matemática, Astronomía, Física y Computación, Universidad Nacional de Córdoba, X5000HUA Córdoba, Argentina}
\affiliation{Instituto de Física Enrique Gaviola, CONICET - UNC, Córdoba, Argentina}

\author{A. Valdés-Hernández}
\affiliation{Instituto de F\'{\i}sica, Universidad Nacional Aut\'{o}noma de M\'{e}xico, Apartado Postal 20-364, Ciudad de M\'{e}xico, M\'{e}xico.}

\author{F. H. Holik}
\affiliation{Universidad Nacional de Hurlingham (UNAHUR), Laboratorio de Investigación y Desarrollo Experimental en Computación (LIDEC), Hurlingham, Argentina}

\author{A. P. Majtey }
\affiliation{Facultad de Matemática, Astronomía, Física y Computación, Universidad Nacional de Córdoba, X5000HUA Córdoba, Argentina}
\affiliation{Instituto de Física Enrique Gaviola, CONICET - UNC, Córdoba, Argentina}

\date{\today}
	
\begin{abstract}

Entanglement in indistinguishable particle systems can be characterized by the impossibility of unambiguously attributing a complete set of physical properties to the individual constituent particles. 
In this work, we introduce a geometric quantifier of the incompatibility of such simultaneous property attribution for pure states of $N$ indistinguishable bosons. Using the Majorana stellar representation, any symmetric multi-qubit state can be expressed as the symmetrization of constituent single-particle states, allowing the associated single-particle properties to be directly linked with the corresponding Majorana stars. This establishes a direct connection between the geometry of the Majorana representation on the Bloch sphere and the entanglement criterion based on the attribution of single-particle properties in systems of identical two-level bosons. We then generalize this geometric approach to higher-dimensional bosons, extending the quantifier from qubits to qudits, while retaining a geometric interpretation of property-attribution incompatibility in terms of Fubini-Study angles.

\end{abstract}
	
\maketitle

\section{Introduction} \label{sec: introduction}

Entanglement in systems of indistinguishable particles is a foundational phenomenon that plays a central role in quantum information science, quantum metrology, and quantum optics \cite{Eckert_2002,tichy_2011_JPB}. 
Unlike systems composed of distinguishable particles, composites of identical bosons or fermions are constrained by the symmetrization postulate. 
This fundamental requirement prevents resorting to  the standard tensor-product partitioning of the Hilbert space to identify physically addressable single-particle subsystems, making the usual notion of local operational partitions inapplicable, and leading to the development of different approaches to study entanglement in indistinguishable-particle systems \cite{Ghirardi_2002,Benatti_2020,killoran_2014,lofranco_2016,lofranco_2018,Wiseman2003,bouvrie_2017_b}. 

One of these approaches is due to Ghirardi, Marinatto, and Weber (GMW), who established a physically motivated property-based criterion for separability \cite{Ghirardi_2002,Ghirardi2004}. 
Within the GMW framework, a pure state of indistinguishable particles is considered non-entangled (separable) if a complete set of physical properties can be unambiguously assigned, with certainty, to each of its individual constituent systems. 
Accordingly, a state is GMW-entangled when such an unambiguous assignment is not possible. 
For fermionic systems, this criterion has led to a comparatively well-developed framework for identifying and quantifying correlations beyond those rooted at the antisymmetry. 
In particular, pure states that can be written as a single Slater determinant are regarded as non-entangled according to the GMW criterion, while states with Slater rank larger than one contain correlations beyond the exchange contribution  \cite{Eckert_2002,pauskaukas_2001,Ghirardi2004}.
Several (fermionic) entanglement measures have subsequently been proposed within this approach, including measures based on the Slater rank, the von Neumann entropy and the linear entropy of the single-particle reduced density matrix, as well as fermionic concurrence \cite{Eckert_2002,manzano_2010,tichy_2011_JPB,majtey_2016}. 
Although the definition and quantification of entanglement for indistinguishable particles remain subject of ongoing debate, these results provide a relatively clear framework for fermionic systems. 
For bosons, however, an analogous quantitative characterization is considerably less established. Recently, qualitative classification schemes have extended this framework to identify different degrees of separability in multipartite bosonic states \cite{Cespedes2025}.

In general, elaborating a quantitative description of non-classical correlations in indistinguishable particle systems introduces conceptual and mathematical subtleties. 
Particularly, when dealing with distinguishable particle systems, entanglement measures are anchored in the operational framework of Local Operations and Classical Communication (LOCC) \cite{amico_2008,horodecki_2009}. 
For indistinguishable parties, however, a universally accepted LOCC resource theory does not exist, the notion of fixed local laboratories is absent, and the class of allowed local operations remains operationally ambiguous. 
Consequently, the requirement of non-increase under LOCC cannot be straightforwardly imported for the entanglement measures.

Rather than imposing an operational structure from distinguishable systems that is not naturally available, an alternative robust approach is to construct a physically grounded quantifier of {\it single-particle property-attribution incompatibility}. Even independently of whether one adopts, or not, the GMW framework to define indistinguishable-particle entanglement, quantifying the degree of mutual incompatibility among $N$ single-particle properties provides  a well-defined figure of merit. 
Such quantity must be a continuous and non-negative functional of the state of the system, and must vanish if and only if all single-particle properties are fully compatible (i.e., mutually identical or exclusive), in the sense described below. 
It also should remain invariant under particle permutations and global single-particle unitary transformations of the form $U^{\otimes N}$.

In this work, we introduce a quantifier with the aforementioned characteristics, valid for pure states of $N$ indistinguishable bosons and based on the pairwise (particle-particle) property-attribution incompatibility. 
Our construction admits a transparent geometrical interpretation, which in the case of two-level bosons exploits the Majorana stellar representation \cite{Majorana1932, Devi2012Majorana,Bengtsson_Zyczkowski_2006}, encoding symmetric multiqubit pure states into constellations of $N$ points (`stars') on the Bloch sphere. 
In this way, a quantitative characterization of GMW entanglement, or equivalently, a characterization of the physical property attributions of the individual parties, is amenable to a geometric visualization.  

The Majorana stellar framework offers significant conceptual and computational advantages over traditional algebraic formulations \cite{Kam2026}. 
Its geometric approach provides a basis-independent representation of quantum states, is manifestly invariant under local unitary symmetries ($SU(2)$), and allows multi-particle correlation structures to be visualized directly through spatial star configurations \cite{Kam2026, Devi2012Majorana}.
However, existing applications of the Majorana framework in the context of entanglement focus primarily on composites of distinguishable parties. 
Non-classical correlations are thus typically characterized via tensor-product partitions, standard LOCC protocols, or qualitative SLOCC classifications based on star degeneracy patterns \cite{Kam2026}. 
In contrast, our work applies the geometric lens directly to composite systems of indistinguishable bosons under the GMW property criterion \cite{Ghirardi_2002,Ghirardi2004,benatti_2014}.  
This establishes a direct link between the physical property-assignment paradigm for identical particles and the constellation geometry on the Bloch sphere, filling a key gap in the quantitative description of non-classical correlations in bosonic systems.
Moreover, the proposed quantifier  naturally extends to higher-dimensional boson systems (qudits instead of qubits), where the Majorana decomposition is no longer valid, and the resulting extension retains a geometric interpretation in terms of Fubini–Study angles.

The article is organized as follows. In Sec. \ref{sec: preliminaries}, we briefly review the GMW separability criterion and the Majorana stellar representation for symmetric $N$-qubit states. 
In Sec. \ref{sec: quantifier}, we formally introduce the quantifier of simultaneous property-attribution incompatibility and discuss its geometric interpretation on the Bloch sphere. We also analyze the Majorana constellations  for representative classes of highly (GMW)-entangled states of 2, 3, and 4 bosonic qubits, detailing the geometric constraints for $N \ge 4$. 
In Sec. \ref{sec: generalizacion}, we discuss the challenges involved in generalizing the quantifier to higher-dimensional bosons and propose an extension to systems of $N$ identical qudits.
Finally, Sec. \ref{sec: Conclusions} presents some final remarks and outlook.

\section{Preliminaries: GMW criterion and Majorana representation}
\label{sec: preliminaries}

In this section, we review the key elements required to establish our geometric quantifier: the GMW property-based criterion for identical bosons, and the Majorana stellar representation for $N$-qubit symmetric states.

\subsection{Ghirardi-Marinatto-Weber (GMW) separability criterion}

For pure states of identical bosons, the usual notion of separability based on assigning a well-defined state to each specific particle cannot be directly applied, since the particles are fundamentally indistinguishable.
Ghirardi, Marinatto, and Weber, instead of relying on artificial particle labels, introduced a property-based criterion for separability \cite{Ghirardi_2002}, characterizing separable (non-entangled) states in terms of the possibility of attributing a complete set of properties to each constituent. 

Consider the two-boson state obtained by symmetrizing the single-particle states $\ket{\psi_1}$ and $\ket{\psi_2}$. 
If these are orthogonal, the corresponding single-particle projectors $P=|\psi_1\rangle\langle\psi_1|$ and $Q=|\psi_2\rangle\langle\psi_2|$ are mutually exclusive ($PQ=0$) so the symmetrized state describes a situation in which, with certainty, one boson possesses all the properties associated with $P$, while the other possesses those associated with $Q$. Although the particles cannot be individually labeled, the two distinct sets of properties can therefore be consistently attributed to the two constituents.
If, instead, $\ket{\psi_1}$ and $\ket{\psi_2}$ are not orthogonal (yet do not represent the same ray in Hilbert space, so $\ket{\psi_1}\neq e^{i\alpha}\ket{\psi_2})$, the two sets of properties are not mutually exclusive. 
As a consequence, there is a  probability $0<p<1$ of finding both bosons in the same single-particle state, so one can no longer assert that one boson possesses the properties associated with \(P\) and the other those associated with \(Q\). 
Thus, according to the GMW criterion, the existence of orthogonal single-particle projectors $P$ and $Q$ such that $\langle \Phi|\mathcal{E}_{P}|\Phi\rangle=\langle \Phi|\mathcal{E}_{Q}|\Phi\rangle=1$, with 
\beq
\mathcal{E}_P = P \otimes (\mathbb I -P) + (\mathbb I-P) \otimes P,
\eeq
(and an analogous $\mathcal{E}_Q$), allows two complete sets of properties to be attributed unambiguously, and with certainty, to the two-boson state $\ket{\Phi}$. 
Clearly, the additional case in which both bosons occupy the same state ($\ket{\Phi}=\ket{\psi}\otimes\ket{\psi}$) should also be regarded as non-entangled (in which case $P=Q$).

In general, assume that an $N$-boson state can be expressed as the symmetrized product of single-particle states $|\psi_i\rangle$,
\begin{equation}
|\Psi\rangle = \mathcal{N}\mathcal{S} \Big( |\psi_{1}\rangle \otimes |\psi_{2}\rangle \otimes \dots \otimes |\psi_{N}\rangle \Big),
\label{eq:symmetric_state}
\end{equation}
where $\mathcal{S}$ is the symmetrization operator and $\mathcal{N}$ a normalization factor.
In line with the GMW approach, each of the single-particle states $\ket{\psi_i}$ can be directly associated to the property of each subsystem.
Then, according to the property-based entanglement criterion, the state in Eq.~(\ref{eq:symmetric_state}) is non-entangled if and only if every pair $\{|\psi_i\rangle,|\psi_j\rangle\}$ satisfies 
\begin{equation}
|\langle\psi_i|\psi_j\rangle|^2 = 1 \quad \text{or} \quad |\langle\psi_i|\psi_j\rangle|^2 = 0.
\label{eq:GMW_condition}
\end{equation}
That is, all underlying single-particle properties are either identical or mutually exclusive. 
Any state (\ref{eq:symmetric_state}) for which $0 < |\langle\psi_i|\psi_j\rangle|^2 < 1$ holds for at least one pair does not admit the attribution of a complete set of well-defined single-particle properties to each of its constituent particles, and is therefore GMW-entangled \cite{Cespedes2025, Ghirardi_2002}. 

Now, for a system of $N$ indistinguishable bosons each one having two orthogonal accessible states (qubits), the Majorana representation guarantees that any composite state admits a decomposition  of the form of the symmetrized product, Eq. (\ref{eq:symmetric_state}) \cite{Devi2012Majorana}. 
In Sec. \ref{sec: quantifier} we focus first on this particular system (two-level bosons) to develop our quantifier, which will then be generalized to composites of $d$-level bosons, with $d>2$.

\subsection{Majorana stellar representation for qubit systems}

In \cite{Majorana1932} Majorana established an isomorphism between the Hilbert space of a $N/2$-spin particle and the symmetric subspace of $N$ qubits. 
This correspondence implies that any pure symmetric state of $N$ qubits admits a unique decomposition (up to a global phase factor) as the symmetrization of a product of $N$ single-qubit states, as expressed in Eq. (\ref{eq:symmetric_state}) \cite{Devi2012Majorana}.

Each constituent single-particle state $|\psi_i\rangle$ in (\ref{eq:symmetric_state}), 
\beq
\ket{\psi_i}=\cos\frac{\vartheta_i}{2}\ket{0}+\sin\frac{\vartheta_i}{2}e^{i\varphi_i}\ket{1},
\eeq
uniquely defines a point (a `star') on the unit sphere $\mathbb{S}^2$, represented by the corresponding Bloch vector $\boldsymbol{r}_i$:
\beq
\boldsymbol{r}_i=(\sin\vartheta_i\cos\varphi_i,\sin\vartheta_i\sin\varphi_i, \cos\vartheta_i).
\eeq
In this picture, an $N$-qubit symmetric state is thus mapped into a set of $N$ stars, or a constellation, on $\mathbb{S}^2$.

Combining the GMW criterion with the Majorana representation yields a simple geometric translation of GMW separability onto the Bloch sphere: identical single-particle states ($|\langle\psi_i|\psi_j\rangle|^2 = 1$) correspond to coinciding Bloch vectors ($\boldsymbol{r}_i = \boldsymbol{r}_j$); orthogonal single-particle states ($|\langle\psi_i|\psi_j\rangle|^2 = 0$) correspond to antipodal Bloch vectors ($\boldsymbol{r}_i = -\boldsymbol{r}_j$).
Consequently, a pure symmetric $N$-qubit state is GMW-separable if and only if all its $N$ Majorana stars  either completely overlap into a single point, or point in antipodal directions. 
Any departure from these 1- or 2-point constellations certifies the superposition of single-particle properties, providing the geometric basis for the quantifier introduced in Sec \ref{sec: quantifier}.

\section{Quantifier of property-attribution incompatibility}\label{sec: quantifier}

To define the quantifier of property-attribution incompatibility in the sense of GMW, we begin by considering the case of two indistinguishable bosons with two degrees of freedom each. As follows from the Majorana decomposition, any symmetric pure state of qubits can be written as
\begin{equation} \label{eq: sym_2}
|\Psi\rangle = \mathcal{N} \Big( |\psi_1\rangle  |\psi_2\rangle+|\psi_2\rangle  |\psi_1\rangle \Big)
\end{equation}
 with $\mathcal{N}$ a normalization constant. 
 As stated above, the state \eqref{eq: sym_2} is not entangled if and only if $\ket{\psi_1}$ and $\ket{\psi_2}$ are either equal (up to a global phase factor) or orthogonal. 
 Thus, in order to capture the GMW entanglement, we look for a real quantity that vanishes if and only if the fidelity $F_{12}=|\langle\psi_1|\psi_2\rangle|^2$ equals 0 or 1. 
 The simplest (polynomial) function of $F_{12}$ satisfying this condition is
\begin{equation} \label{eq: medida_2particulas}
E_2(\Psi) = 4 \, F_{12} \left( 1 - F_{12} \right),
\end{equation}
where the factor $4$ has been introduced in order to bound  $E_2\in[0,1]$.

The quantity (\ref{eq: medida_2particulas}) can be straightforwardly extended to any bosonic pure state of $N$ qubits in the form (\ref{eq:symmetric_state})
by taking into account the contribution of each (pairwise) fidelity $F_{ij}=|\langle\psi_i|\psi_j\rangle|^2$. 
Thus, we define 
\begin{equation}\label{eq: E_N con F}
E_N(\Psi) = \frac{4}{\binom{N}{2}} \sum_{i < j}^N F_{ij} ( 1 - F_{ij}),
\end{equation}
where the binomial factor $\binom{N}{2} = \frac{N(N-1)}{2}$, counting the number of different pairs in the set $\{\ket{\psi_i}\}$, has been introduced for normalization purposes.
The quantity (\ref{eq: E_N con F}) is the simplest polynomial function of pairwise fidelities that vanishes if and only if the qubits occupy equal or exclusive states.
Further, $E_N$ is trivially invariant under permutation of the parties, and under unitaries $U^{\otimes N}$.

$E_N$ is also amenable of a geometric interpretation by expressing the pairwise fidelity as
\begin{eqnarray}\label{eq: fid}
F_{ij}&=&|\langle\psi_i|\psi_j\rangle|^2=\frac{1+\boldsymbol{r}_i\cdot\boldsymbol{r}_j}{2}\nonumber\\
&=&\frac{1+\cos\theta_{ij}}{2}=\cos^2\Big(\frac{\theta_{ij}}{2}\Big),
\end{eqnarray}
where $\theta_{ij}$ stands for the relative angle between the Bloch vectors $\boldsymbol{r}_i$ and $\boldsymbol{r}_j$.
Each (pairwise) contribution to (\ref{eq: E_N con F}) thus rewrites as 
\beq\label{eq: fij}
4F_{ij}(1-F_{ij})=\sin^2 \theta_{ij}.
\eeq
It takes its lowest (null) value whenever the angles are either $\theta_{ij} = 0$ or $\theta_{ij} = \pi$ ---which correspond to equal or orthogonal properties, respectively---,  
and reaches its maximum (unit) value provided $\theta_{ij} = \pi/2$.

On the other hand, the (normalized) chord distance between $\boldsymbol{r}_i$ and $\boldsymbol{r}_j$ reads 
\beq
\tilde d_{ij}=\frac{1}{2}|\boldsymbol{r}_i-\boldsymbol{r}_j|=\frac{1}{2}\sqrt{2(1-\boldsymbol{r}_i\cdot\boldsymbol{r}_j)}=\sin\frac{\theta_{ij}}{2}
\eeq
and relates to the fidelity (\ref{eq: fid}) according to 
\beq
\tilde d_{ij}^2=1-F_{ij},
\eeq
which is the distinguishability between single particle states.
Therefore, Eq. (\ref{eq: E_N con F}) becomes
\begin{equation}\label{eq: En con dij}
E_N(\Psi) = \frac{8}{N(N-1)} \sum_{i < j}^N  \tilde{d}^2_{ij}(1-\tilde{d}^2_{ij}).
\end{equation}
Now, the triangle formed by the vectors $\boldsymbol{r}_i$, $\boldsymbol{r}_j$, and $\boldsymbol{r}_i-\boldsymbol{r}_j$ (with side lengths $1$, $1$, and $2\tilde{d}_{ij}$, respectively) has an area given by $\mathcal{A}_{ij} = \tilde{d}_{ij}\sqrt{1-\tilde{d}_{ij}^2}$. 
Consequently, $E_N(\Psi)$ in Eq.~(\ref{eq: En con dij}) is directly proportional to the sum of these squared triangular areas.
This  makes transparent that $E_N$ vanishes for collinear ($\tilde{d}_{ij}=0$) or antipodal ($\tilde{d}_{ij}=1$) vectors, and reaches its highest value ($E_N=1$) when all the vectors are mutually orthogonal, so $\mathcal{A}_{ij}=1/2$ (a condition that, as will be seen below, can only hold for $N\leq 3$).


In order to identify the states with higher values of $E_N$, we first determine its precise maximal value as follows. 
Substituting $\sin^2\theta_{ij}=1-(\boldsymbol{r}_i\cdot \boldsymbol{r}_j)^2$ into Eq. (\ref{eq: fij}) and resorting to (\ref{eq: E_N con F}) we obtain
\begin{equation}\label{eq: EN rirj}
E_N(\Psi) = 1-\frac{2}{N(N-1)} \sum_{i < j}^N (\boldsymbol{r}_i\cdot \boldsymbol{r}_j)^2,
\end{equation}
so maximization of $E_N$ requires minimizing the sum $\sum_{i < j}^N (\boldsymbol{r}_i\cdot \boldsymbol{r}_j)^2$. To this aim we consider the second-moment matrix
\beq
M=\sum_i\boldsymbol{r}_i\boldsymbol{r}^{T}_i=\sum_i\begin{pmatrix}
x^2_i & x_iy_i & x_iz_i\\
x_iy_i & y^2_i & y_iz_i\\
x_iz_i&y_iz_i&z^2_i
\end{pmatrix}
\eeq
that satisfies:
\begin{subequations}
\begin{eqnarray}
\text{Tr} \,M&=&N,\\
\text{Tr} \,M^2&=&\sum_{i,j}(\boldsymbol{r}_i\cdot \boldsymbol{r}_j)^2=N+2\sum_{i<j}(\boldsymbol{r}_i\cdot \boldsymbol{r}_j)^2.\label{eq: trM2}
\end{eqnarray}
\end{subequations}
Consequently, the problem of maximizing $E_N$ reduces to the minimization of $\text{Tr} \,M^2$ subject to the constrain $\text{Tr} \,M=N$. Let $\{\mu_1,\mu_2,\mu_3\}$ be the eigenvalues of $M$; then the solution of the optimization problem is 
\beq\label{eq: opt}
\mu_1=\mu_2=\mu_3=\mu=\frac{N}{3},
\eeq
provided that $N\geq 3$ (since $M$ is a $3\times3$ matrix given by the sum of $N$ rank-one matrices, it holds that $
\operatorname{rank}(M)\leq\min\{N,3\}$. Thus, for the three eigenvalues $\mu_k$ to be equal to the nonzero value $N/3$, $M$ must have rank $3$, which is possible only if $N\geq3$).

Eq. (\ref{eq: opt}) gives, from Eq. (\ref{eq: trM2}),
\begin{equation}
\Big[\sum_{i< j}^N (\boldsymbol{r}_i\cdot \boldsymbol{r}_j)^2\Big]_{\min}=\frac{1}{2}\Big(\frac{N^2}{3}-N\Big) \quad{(N\geq 3),}
\end{equation}
leading finally to
\begin{equation}\label{eq: EN max}
(E_N)_{\max} = \frac{2}{3}\frac{N}{N-1}\leq 1 \quad (N\geq 3).
\end{equation}

Explicitly, the optimal single-particle Bloch vector configurations $\{\boldsymbol{r}_i = (x_i, y_i, z_i)^T\}_{i=1}^N$ that achieve this maximum are obtained by solving the constraint $M = \frac{N}{3}\mathbb{I}_3$, which translates into the general system of equations:
\begin{eqnarray}
\sum_{i=1}^N x_i^2 &=& \sum_{i=1}^N y_i^2 = \sum_{i=1}^N z_i^2 = \frac{N}{3}, \nonumber\\
\sum_{i=1}^N x_i y_i &=& \sum_{i=1}^N x_i z_i = \sum_{i=1}^N y_i z_i = 0,
\end{eqnarray}
subject to $x_i^2 + y_i^2 + z_i^2 = 1$ for all $i \in \{1, \dots, N\}$.

According to Eq.~(\ref{eq: EN max}), the case $N=3$ is the only one for which the maximum value of $E_N$ reaches $1$ (besides, of course, the $N=2$ case, as follows from Eq. (\ref{eq: medida_2particulas})). 
As $N$ increases, $(E_N)_{\max}$ decreases and approaches the asymptotic value $2/3$.

\begin{figure}[htbp]
    \centering    \includegraphics[width=0.5\columnwidth]{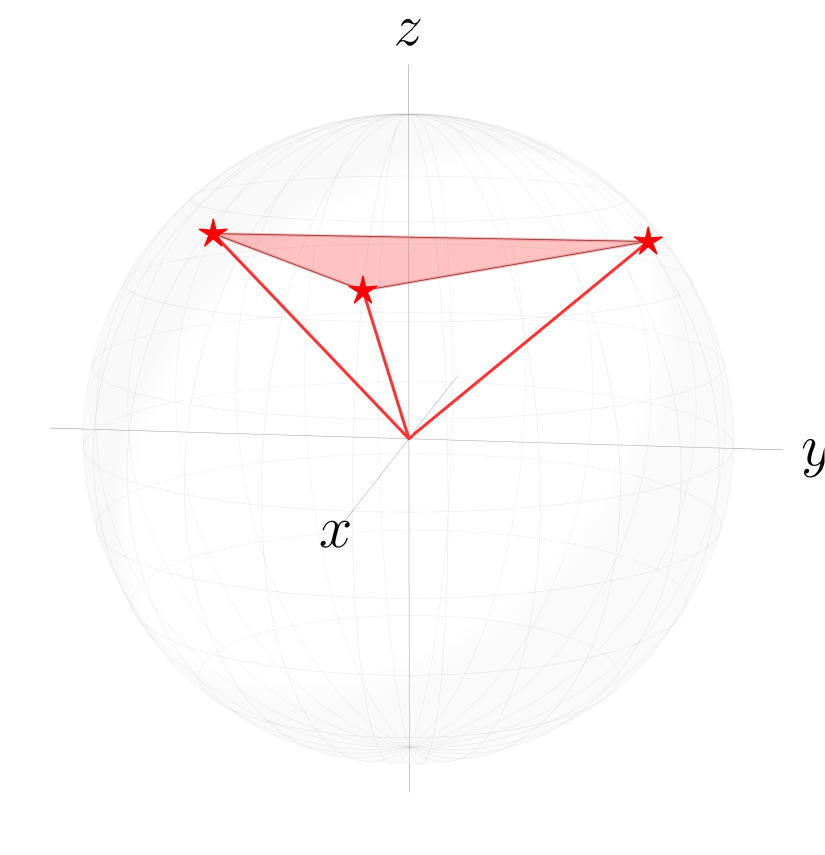}
\includegraphics[width=\columnwidth]{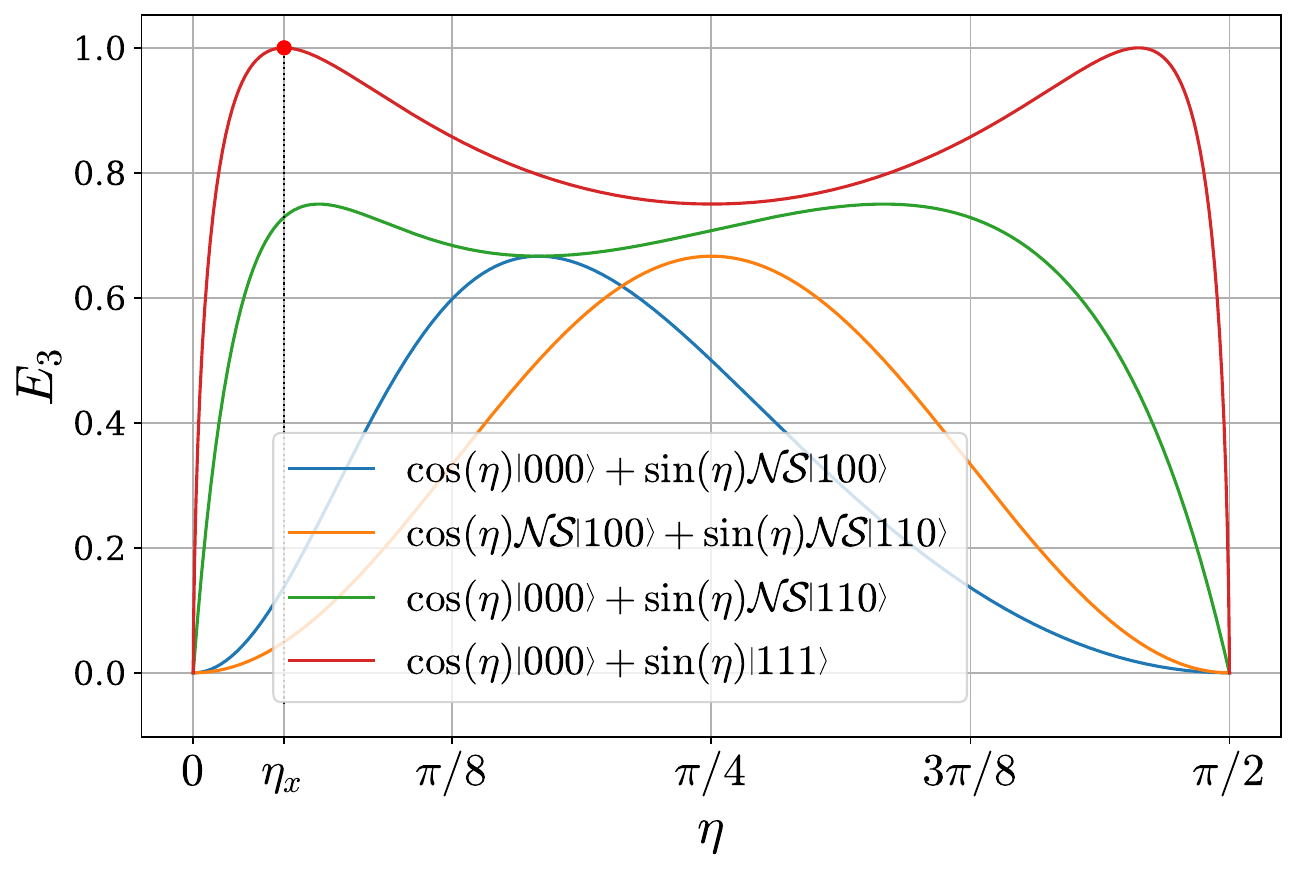}
    \caption{\justifying Top panel: Majorana constellation for the GHZ-type state with maximum $E_3$, corresponding to the red dot in the bottom panel ($ \eta_x \approx 0.13782$). All other maximizing states are obtained by the orbit under rigid rotations of that state.
    Bottom panel: $E_{3}$ as a function of $\eta$ for different one-parameter families of three-qubit states.}
    \label{fig: entanglement_3p}
\end{figure}

For $N=3$, the maximum $(E_N)_{\max}=1$ is reached for states $\ket{\Psi} = \mathcal{N}\mathcal{S}\left(\ket{\psi_1} \otimes \ket{\psi_2} \otimes \ket{\psi_3}\right)$ such that $\{\boldsymbol r_1,\boldsymbol r_2,\boldsymbol r_3\}$ form an orthogonal triad, as shown in the top panel of Figure \ref{fig: entanglement_3p}. 
The bottom panel shows $E_3$ for 4 families of three-qubit states, as they vary with the real parameter $\eta$. 
 The point highlighted in the red curve corresponds to a GHZ-type state with maximum $E_3$. 
This state decomposes as a symmetrization of three completely non-orthogonal single-particle states $\{\ket{\psi_i}\}$, represented by the (orthogonal) Bloch vectors in the top panel of Fig. \ref{fig: entanglement_3p}. 
An analogous situation occurs for the second maximum in the red curve, simply replacing $\eta_x\rightarrow (\pi/2)-\eta_x$, and reflecting the constellation with respect to the equator of the sphere.

\begin{figure}[htbp] 

    \centering
    \begin{subfigure}[b]{0.45\linewidth}
        \centering
        \includegraphics[width=\linewidth]{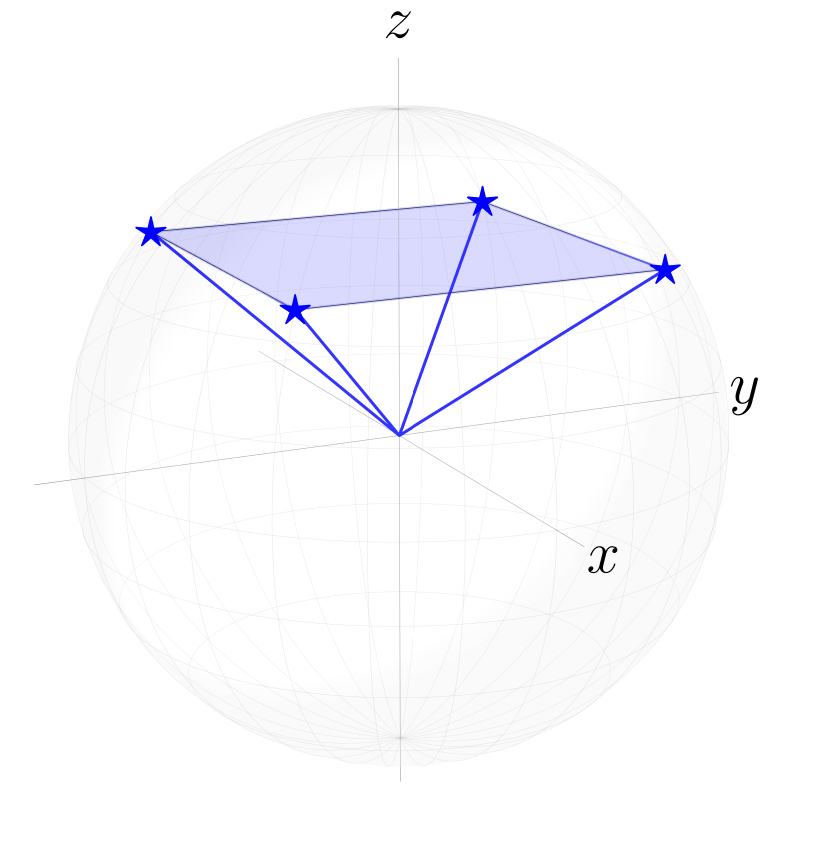} 
        \caption{}
        \label{fig:esfera_a}
    \end{subfigure}
    \hfill 
    \begin{subfigure}[b]{0.45\linewidth}
        \centering
        \includegraphics[width=\linewidth]{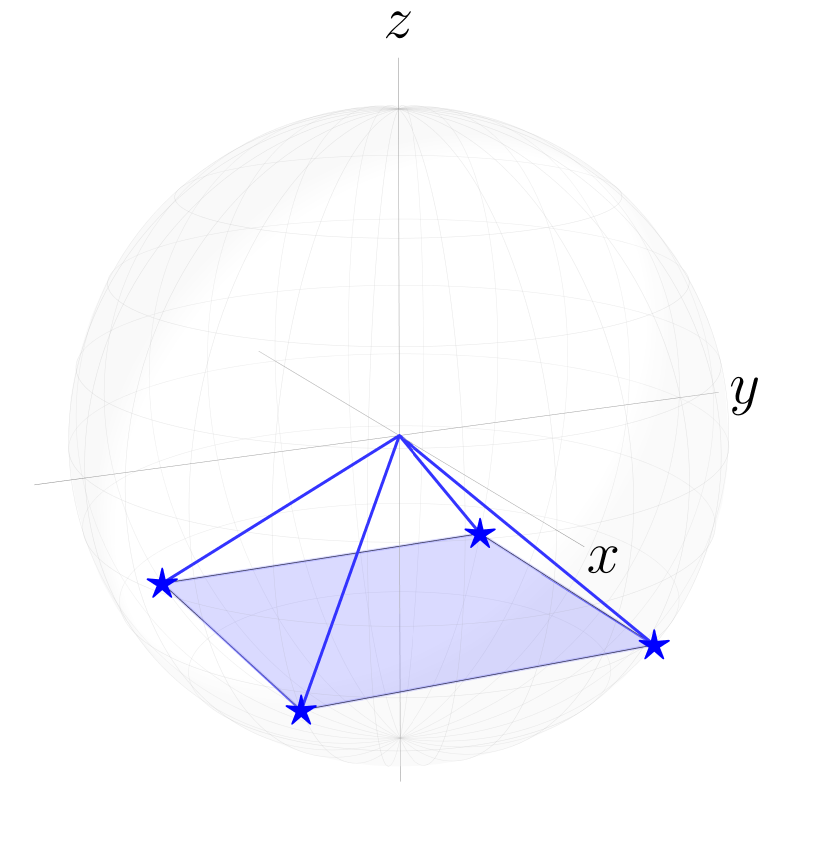}
        \caption{}
        \label{fig:esfera_b}
    \end{subfigure}
 \vspace{0.4cm} 
    \begin{subfigure}[b]{0.45\linewidth}
        \centering
        \includegraphics[width=\linewidth]{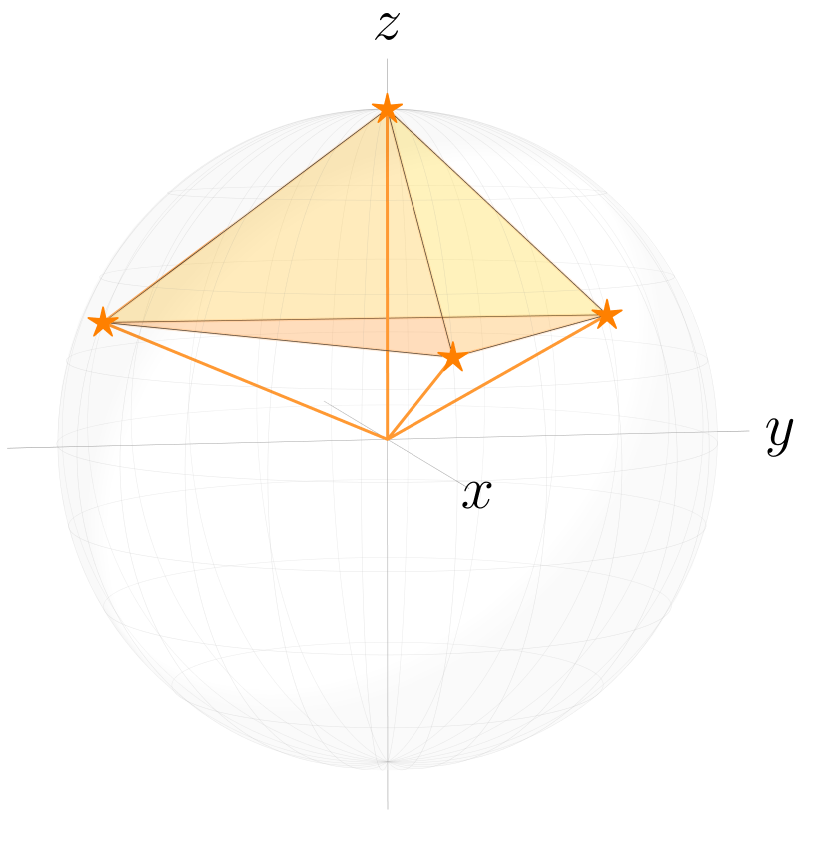}
        \caption{}
        \label{fig:esfera_c}
    \end{subfigure}
    \hfill
    \begin{subfigure}[b]{0.45\linewidth}
        \centering
        \includegraphics[width=\linewidth]{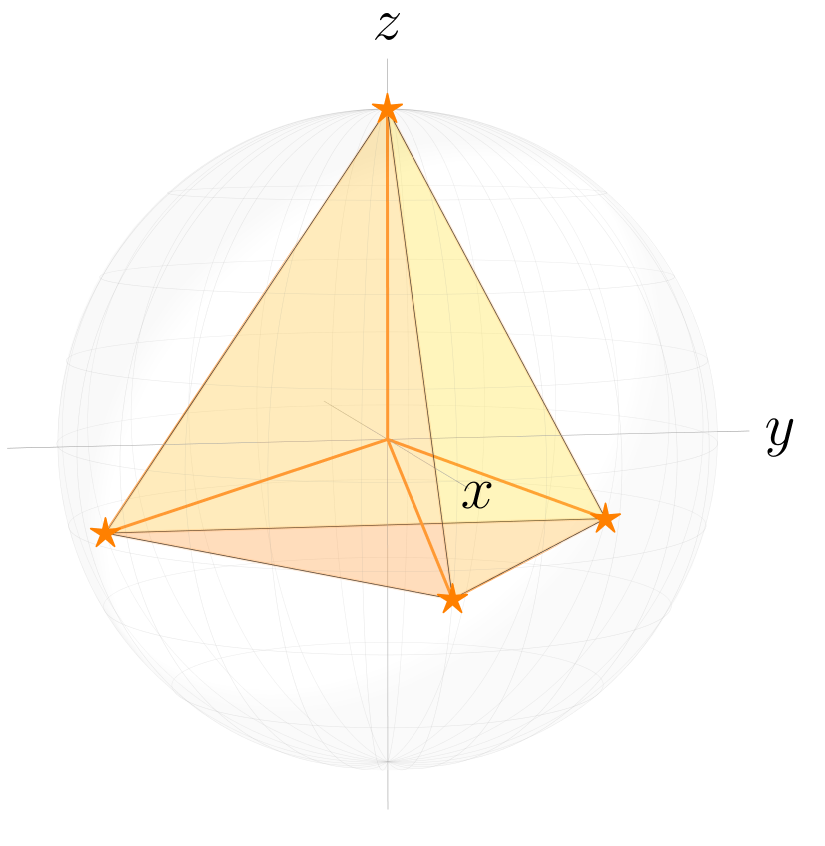}
        \caption{}
        \label{fig:esfera_d}
    \end{subfigure}
\vspace{0.5cm} 
    \begin{subfigure}[b]{\linewidth}
        \centering
   \includegraphics[width=0.99\linewidth]{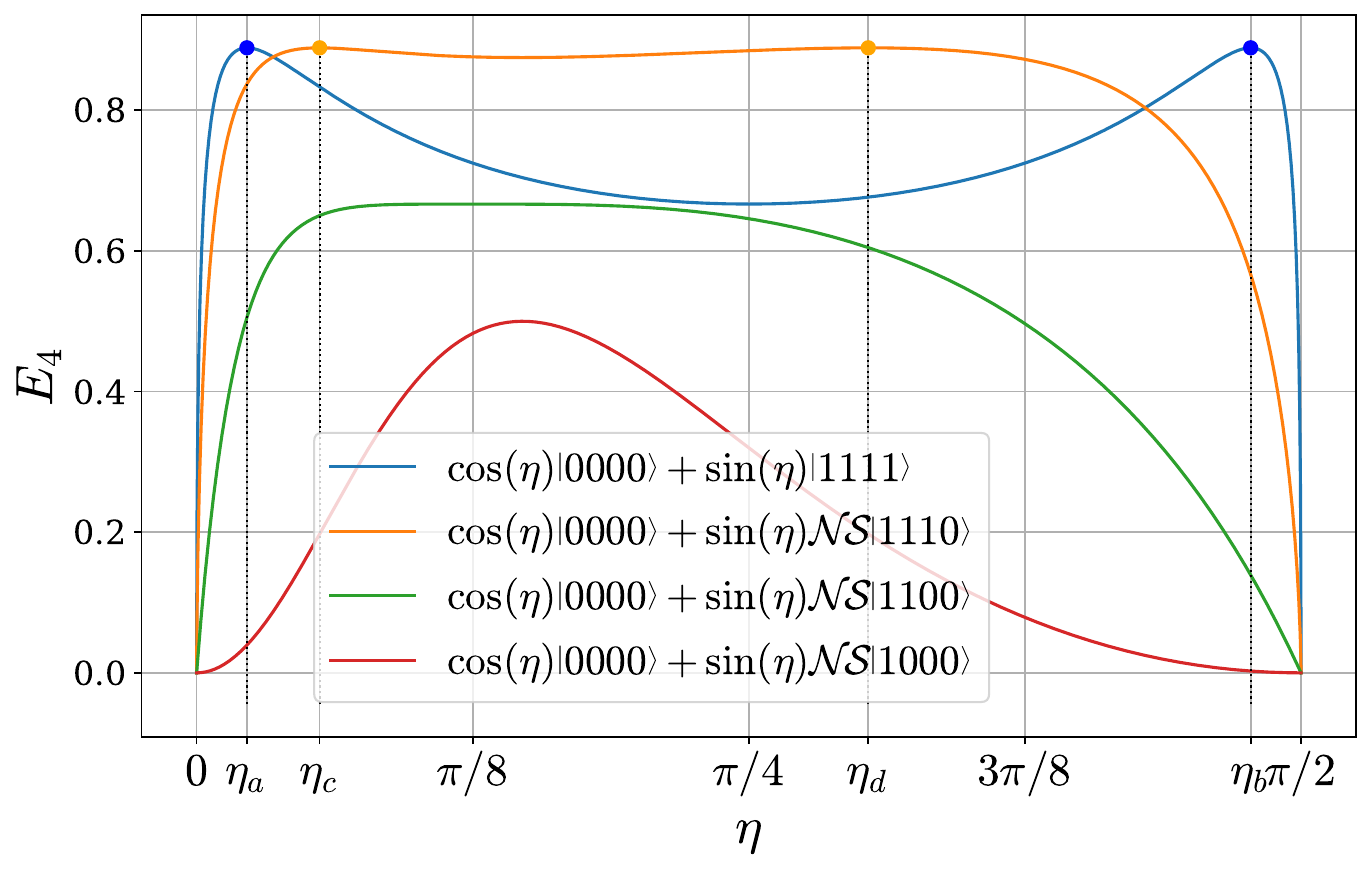}
        \caption{}
        \label{fig:curvas_E4}
    \end{subfigure}

    \caption{\justifying Panels
    (a)--(d): Constellations of four-qubit states reaching the maximum value of $E_4$. Panels (a) and (b) correspond to the blue state family at $\eta_a\approx 0.0715$ and $\eta_b = \pi/2 - \eta_a$, respectively, while (c) and (d) correspond to the orange state family at $\eta_c \approx 0.1750$ and $\eta_d \approx 0.9552$, respectively.  Panel (e): $E_4$ as a function of $\eta$ for different one-parameter families of four-qubit states, with dots indicating the states depicted 
    in panels (a)--(d).}
    \label{fig: entanglement_4p}
\end{figure}

Whereas for $N=2$ and $N=3$ the distribution of the Majorana stars corresponding to 
states with maximum $E_N$ is straightforward (given by Bloch vectors that conform an orthogonal dyad and triad, respectively), for $N=4$ the complete mutual orthogonality among all $\boldsymbol{r}_i$ ($i=1,\dots,4$) becomes geometrically impossible on $\mathbb{S}^2$.
Interestingly, the optimization of $E_4$ reveals a geometric degeneracy, since the  maximum $(E_4)_{\text{max}} = 8/9$ is realized by three $SU(2)$-inequivalent star constellations,
represented in Figure \ref{fig: entanglement_4p}, and which define: a square on a constant-latitude plane (panels (\ref{fig:esfera_a}) and (\ref{fig:esfera_b})), a non-regular tetrahedron (panel (\ref{fig:esfera_c})), and a regular tetrahedron (panel (\ref{fig:esfera_d})).
These constellations represent specific four-qubit states highlighted with color dots on the curves of $E_4$ shown in the panel (e) of Fig. \ref{fig: entanglement_4p}. 
Along the orange curve, 
the first maximum corresponds to the non-regular tetrahedron, yielding three pairs with $\boldsymbol r_i\cdot\boldsymbol r_j=-1/3$, corresponding to the Bloch vectors pointing to the triangular base vertices, and three pairs with $\boldsymbol r_i\cdot\boldsymbol r_j=1/3$, involving the Bloch vector pointing to the apex and the remaining three. 
The second maximum corresponds to the  regular tetrahedron, where all six pairwise dot products satisfy $\boldsymbol{r}_i \cdot \boldsymbol{r}_j = -1/3$. 

Along the blue curve, the first maximum corresponds to a four-qubit GHZ-type state, and the four Majorana stars form a square on a constant-latitude plane ($\theta = \arccos(1/\sqrt{3})$), yielding four pairs with $\boldsymbol{r}_i \cdot \boldsymbol{r}_j = 1/3$ and two   pairs with $\boldsymbol{r}_i \cdot \boldsymbol{r}_j = -1/3$. 
The second maximum along the blue curve is the symmetric counterpart of the first one, with $\eta_b = (\pi/2)-\eta_a $. 
Consequently, as governed by Eq. (\ref{eq: EN rirj}), these geometrically distinct constellations share the same collective sum of squared inner products, all attaining the maximum $E_4 = 8/9$. This degeneracy reflects the fact that the isotropy condition $M = \frac{N}{3}\mathbb{I}_3$ does not uniquely determine the stellar constellation.

%


\section{Generalization for higher dimensional systems} \label{sec: generalizacion}

When extending the analysis to systems with higher degrees of freedom, so that the single-boson Hilbert space is $d$-dimensional with $d>2$, the underlying entanglement structure in the GMW framework undergoes a fundamental transformation \cite{Cespedes2025}.
In this case there are some states for which no property can be attributed to any of the particles, which  is coincident with the fact that for $d \ge 3$ there exist symmetric multiqudit states that no longer admit a unique factorization into symmetrized tensor products of individual single-particle states \cite{Chryssomalakos2021}, and consequently the  conventional Majorana stellar representation loses its direct applicability. 

For $N$-qudit bosonic states that allow a representation of the form (\ref{eq:symmetric_state}), $E_N$ can be directly computed from Eq. (\ref{eq: E_N con F}). 
In such case, of course, the interpretation in terms of Bloch vectors is no longer valid, yet the expression (\ref{eq: E_N con F}) is still amenable to a geometric interpretation, as will be seen below. 
States that do not admit a decomposition of the form (\ref{eq:symmetric_state}) are necessarily GMW entangled (the simultaneous property attribution does not hold), yet the quantifier (\ref{eq: E_N con F}) loses its applicability (the set $\{\ket{\psi_i}\}$ does not exist).
In what follows we propose a generalization of $E_N$ that holds in general, whether the state admits a symmetrized decomposition or not. 

Given an arbitrary $N$-particle state  $\ket{\Phi}$, the generalization is defined in several steps. First, define the set $S_{0}$ formed by all states $\ket{\Psi}$ of the form \eqref{eq:symmetric_state}. Next, given $|\Phi\rangle$, define $F_{\Phi}^{\text{max}} = \max \{|\langle\Psi|\Phi\rangle |^{2}\,\,\Big|\,\, |\Psi\rangle\in S_{0}\}$. 
Possibly, there might be more than one vector in $S_0$ for which the maximum is attached. Thus, define
\begin{equation}
S_{\Phi}=\{ |\Psi\rangle \in S_{0}\,\,\Big|\,\,|\langle\Psi|\Phi\rangle|^{2}=F_{\Phi}^{\text{max}}\}.
\end{equation}  
This set contains the symmetrized states (satisfying Eq. \eqref{eq:symmetric_state}) that are closest to the state of interest. The quantity $E_N$ in Eq. \eqref{eq: E_N con F} can thus be applied to each element of $S_{\Phi}$, resulting in different scalar values. Define $E_{N,\Phi}^{\text{min}}= \min\{E_N(\Psi)\,\,\Big|\,\,|\Psi\rangle\in S_{\Phi}\}$. We are interested in the states in $S_{\Phi}$ that minimize the value of $E_N$. 
Those are the symmetrized states that, along with having maximal fidelity with $|\Phi\rangle$, minimize the value of $E_N$ (i.e., $E_{N} (\Psi)=E_{N,\Phi}^{\min}$). They form the set:
\begin{equation}
s_{\Phi} = \{|\Psi\rangle \in S_{\Phi} \,\,\Big|\,\, E_N(\Psi) =E_{N,\Phi}^{\text{min}} \}.  
\end{equation}
The generalized quantifier is then defined as
\begin{equation}\label{eq: cuantificador_generalizado}
 \mathbb E_N(\Phi) = \left( 1- F^{\max}_{\Phi} \right) + E^{\min}_{N,\Phi}.    
\end{equation} 

The first term in Eq. (\ref{eq: cuantificador_generalizado}) contributes whenever $\ket{\Phi}$ cannot be written as a symmetrization, and the second term measures the pairwise property-attribution incompatibility of the particles of the optimal symmetrized states in $s_{\Phi}$  
(i.e., those states of the form \eqref{eq:symmetric_state} which are closest to $|\Phi\rangle$ and minimize $E_{N}$).
The additive form in Eq. (\ref{eq: cuantificador_generalizado}) provides the simplest extension that incorporates these two contributions (departure from the set of symmetrized states and the minimal incompatibility of property attribution of the closest symmetrized states). This choice is not intended to be unique, but has the desirable property of reducing to $E_N$ whenever $F_\Phi^{\max}=1$. Notice that, unlike $E_N$, the generalized quantifier is not normalized to unity.

Furthermore, $\mathbb E_{N}(\Phi)$ vanishes if and only if  $|\Phi\rangle$ decomposes as a symmetrized state satisfying both $|\Phi\rangle=\mathcal N\mathcal S(|\psi_1\rangle \otimes |\psi_2\rangle \otimes \dots \otimes |\psi_N\rangle)$ ($F^{\max}_{\Phi}=1$) and all the $\{\psi_i\}$ are equal or pairwise orthogonal   ($E_{N,\Phi}^{\min}=0$). In other words, $\mathbb E_N$ vanishes iff the $N$ qudits possess well-defined orthogonal properties.

This generalized quantifier also admits a simple geometric interpretation in terms of the Fubini--Study metric \cite{Bengtsson_Zyczkowski_2006}.
For pure states, the Fubini--Study angle between $|\Phi\rangle$ and $|\Psi\rangle$ is defined as 
\beq
\gamma(\Phi,\Psi)=\arccos \sqrt{F(\Phi,\Psi)}=\arccos |\langle\Phi|\Psi\rangle|.
\eeq
Since $F^{\max}_{\Phi}$ is the maximum fidelity between $|\Phi\rangle$ and a state $\ket{\Psi}$ admitting a symmetrized product decomposition (\ref{eq:symmetric_state}), the first term in Eq.~\ref{eq: cuantificador_generalizado} can be written as
\begin{equation}
1-F^{\max}_{\Phi}
=\sin^2\gamma_{\min},
\end{equation}
where $\gamma_{\min}$ is the shortest Fubini--Study angle between $|\Phi\rangle$ and the set $S_0$.
A similar geometric interpretation applies to the incompatibility contribution $E_{N,\Psi}^{\min}$. 
For two single-particle states $|\psi_i\rangle$ and $|\psi_j\rangle$, we have $\cos\gamma_{ij}=\sqrt{F_{ij}}$, hence
\begin{equation}\label{Fub}
4F_{ij}(1-F_{ij})
=\sin^2\left(2\gamma_{ij}\right).
\end{equation}
For qubit systems, the Fubini-Study angle $\gamma_{ij}$ becomes $\gamma_{ij}=\theta_{ij}/2$ (with $\theta_{ij}$ the relative angle between the Bloch vectors $\boldsymbol{r}_i$ and $\boldsymbol{r}_j$), so (\ref{Fub}) reduces to $\sin^2\theta_{ij}$, recovering the geometric interpretation discussed above on the Bloch sphere. 
For $d>2$, the same construction naturally extends to the projective space $\mathbb{CP}^{d-1}$ of pure states of qudits, where the pairwise terms in Eq.~(\ref{Fub})  quantify the Fubini--Study angular incompatibility between the states $|\psi_i\rangle$ and $|\psi_j\rangle$.
Thus, $\mathbb{E}_N(\Phi)$ combines two geometric aspects: the distance of the state of interest $\ket{\Phi}$ from the set of symmetrized states $\mathcal{N}\mathcal{S} ( |\psi_{1}\rangle \otimes |\psi_{2}\rangle \otimes \dots \otimes |\psi_{N}\rangle)$, and all the pairwise angular incompatibilities between the corresponding constituent states $\{\ket{\psi_i}\}$.

\begin{figure}[htbp]
    \centering    
\includegraphics[width=\columnwidth]{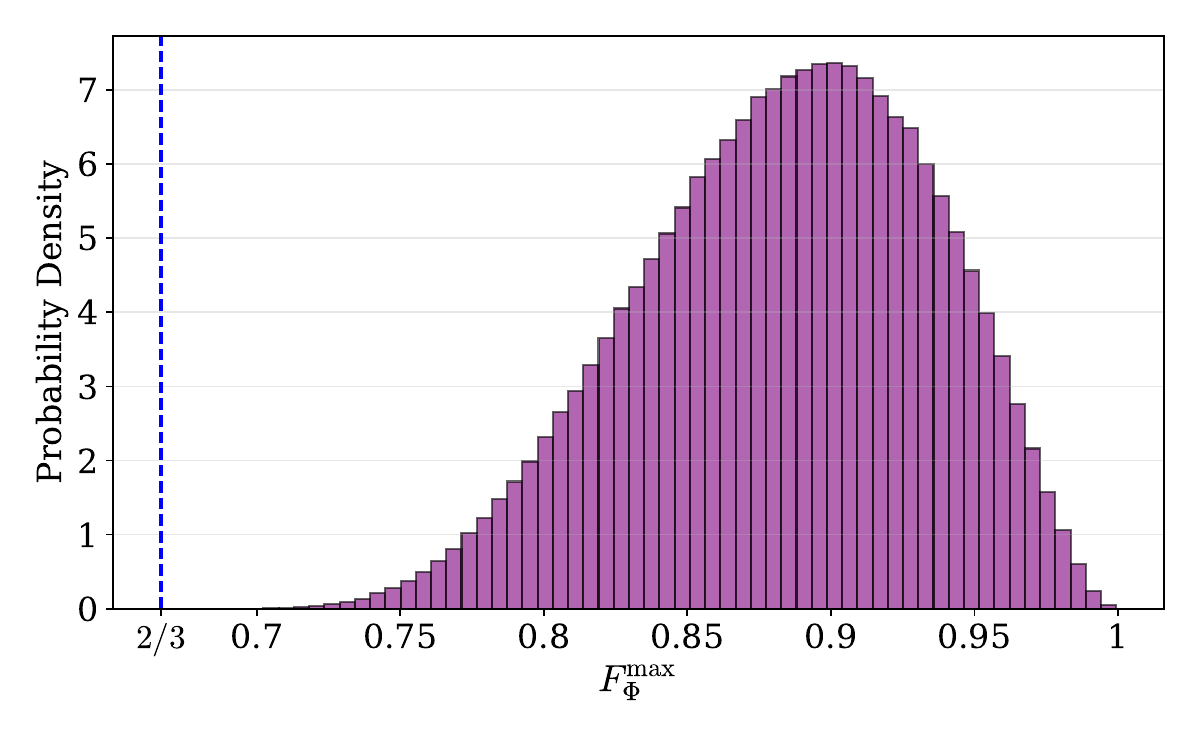}
    \caption{\justifying Probability density distribution of the maximum fidelity $F^{\max}_{\Phi}$ for a 3 qutrit system  ($N=3, d=3$). Histogram obtained from a sampling of Haar-random pure symmetric states in the 10-dimensional Hilbert space.}
    \label{fig: fidelity}
\end{figure}

To illustrate $\mathbb E_N$, we consider three-qutrit states ($N=d=3$). We evaluate the distribution of $F^{\max}_{\Phi}$ by analyzing a random sample of $10^6$ pure states $\vert{}\Phi\rangle$, generated according to the Haar measure \cite{Zyczkowski1998}, which defines a natural uniform distribution over the unit sphere of the 10-dimensional symmetric Hilbert space.
In practice, this is done by constructing complex vectors $\boldsymbol{c} \in \mathbb{C}^{10}$ whose real and imaginary components are independent standard Gaussian variables, to generate the states $|\Phi\rangle = \sum_{j=1}^{10} (c_j/|\boldsymbol{c}|) |\phi_j\rangle $, with $\{|\phi_j\rangle\}$  an orthonormal Fock basis.
For each sampled state $\vert{}\Phi\rangle$, the maximum fidelity $F^{\max}_{\Phi}$ is numerically evaluated by performing a non-linear optimization over all single-particle states $\{|\psi_i\rangle\}$ forming the symmetrized state $\mathcal{N}\mathcal{S}(|\psi_1\rangle \otimes|\psi_2\rangle\otimes|\psi_3\rangle$). 
The resulting probability density distribution of $F^{\max}_{\Phi}$ is presented in Fig. (\ref{fig: fidelity}). No state with $F_\Phi^{\max}<2/3$ was found in our sample of $10^6$ Haar-random states.
The density of random states concentrates predominantly around $F^{\max}_{\Phi} \approx 0.9$,  illustrating that the states that reach the minimal ($F^{\max}_{\Phi} = 1$) and the maximal ($ F^{\max}_{\Phi} = 2/3$) distance observed in our numerical analysis constitute extremal cases that are rarely encountered under random sampling.

In particular, consider the three-qutrit state $\ket{\Phi} = \frac{1}{\sqrt{3}} (\ket{000} + \ket{111} + \ket{222})$.
It cannot be written as a symmetrization of single-particle states \cite{Cespedes2025}, and has $F^{\max}_{\Phi}=2/3$, corresponding to the smallest value observed in our numerical analysis, so the first term in Eq. (\ref{eq: cuantificador_generalizado}) attains its maximal value.   
The optimal state is a symmetrization of a triad $ \left\{\ket{\psi_1},\ket{\psi_2},\ket{\psi_3}  \right\}\ $ where all the $\ket{\psi_i}$ are orthogonal to each other: 
\begin{equation}
\begin{split}
\ket{\psi_1} &= \frac{1}{\sqrt{3}}
\left( \ket{0} + \ket{1} + \ket{2} \right),\\
\ket{\psi_2} &= \frac{1}{\sqrt{3}}
\left( \ket{0} + e^{i \frac{2 \pi }{3}} \ket{1} + e^{i \frac{4 \pi }{3}}\ket{2} \right),\\
\ket{\psi_3} &= \frac{1}{\sqrt{3}}
\left( \ket{0} + e^{i \frac{4 \pi }{3}}\ket{1} + e^{i \frac{2 \pi }{3}}\ket{2} \right).
\end{split}
\end{equation}
Therefore, the second term in \eqref{eq: cuantificador_generalizado} vanishes, and $\mathbb E_N$ reduces to $\mathbb E_N=1-F^{\max}_{\Phi}=1/3$. 

Another example of a  state that cannot be written as a symmetrization of single particle states is $ \ket{\Phi} = \frac{1}{\sqrt{9}} ( \mathcal{S} \ket{001} +\mathcal{S} \ket{112} + \mathcal{S} \ket{220} )$. In this case we also have $F^{\max}_{\Phi}=2/3$, but now the optimal state is a symmetrization of a triad $\{\ket{\psi_1},\ket{\psi_2},\ket{\psi_3}\}$ in which no $\ket{\psi_i}$ is orthogonal to another one:  
\begin{equation}
\begin{split}
    |\psi_1\rangle &= a |0\rangle + b e^{-i \frac{\pi}{3}} |1\rangle + c e^{-i \frac{2\pi}{3}} |2\rangle,
    \\
    |\psi_2\rangle &= b |0\rangle + c e^{-i \frac{\pi}{3}} |1\rangle + a e^{i \frac{\pi}{3}} |2\rangle,
    \\
    |\psi_3\rangle &= c |0\rangle + a e^{i \frac{2\pi}{3}} |1\rangle + b e^{i \frac{\pi}{3}} |2\rangle,
\end{split}
\end{equation}
with $a \approx 0.1418$, $b \approx 0.7673$, and $c \approx 0.6255$.
In this case we get $\mathbb E_N(\Phi) = \frac{1}{3} + 0.75 $.

As a final example we consider a state with a more typical maximal fidelity, according to Fig. \ref{fig: fidelity}. 
It reads
$ \ket{\Phi} = \frac{2}{3} \ket{000} + \frac{2}{3} \ket{111} + \frac{1}{3} \ket{222}$. 
Numerical optimization over the set of symmetrized product states yields a maximum fidelity $F^{\max}_{\Phi} \approx 0.88$, corresponding to a distance contribution $1 - F^{\max}_{\Phi} \approx 0.12$. The closest symmetrized state is given by the constituent single-particle states
\begin{equation}
\begin{aligned}
\ket{\psi_1} &= \frac{1}{\sqrt{3}} \left( \ket{0} + \ket{1} \right), \\
\ket{\psi_2} &= \frac{1}{\sqrt{3}} \left( \ket{0} + e^{-i \frac{2\pi}{3}} \ket{1} \right), \\
\ket{\psi_3} &= \frac{1}{\sqrt{3}} \left( \ket{0} + e^{i \frac{2\pi}{3}} \ket{1} \right).
\end{aligned}
\label{eq:states_example}
\end{equation}
Evaluating the pairwise angular incompatibility among these constituent states gives $E_{N,\Phi}^{\min} = 0.75$. Combining both contributions, the total generalized incompatibility measure for this state is $\mathbb{E}_N(\Phi) = (1 - F^{\max}_{\Phi}) + E_{N,{\Phi}}^{\min} \approx 0.12 + 0.75$.


\section{Final remarks}\label{sec: Conclusions}

We have established a geometric framework to quantify simultaneous property-attribution incompatibility  in indistinguishable boson systems, 
which in addition certifies non-classical correlations among identical bosons 
consistently with the Ghirardi-Marinatto-Weber (GMW) property-assignment paradigm. 
Rather than importing the operational structure from distinguishable systems 
that is not naturally available, our approach provides a physically grounded 
measure of pairwise (boson-boson) incompatibility, in the sense that a complete set of physical properties cannot be assigned simultaneously to (at least) a pair of bosons.

For systems of $N$ two-level bosons, this quantity is explicitly realized by $E_N(\Psi)$.
By exploiting the Majorana stellar representation for symmetric multiqubit states,  $E_N(\Psi)$ admits a geometric representation of the physical attribution of single-qubit properties directly as constellations on the Bloch sphere. 
In particular, $E_N(\Psi)$ is proportional to the mean squared area of the triangles spanned by pairs of Majorana stars from the center of the sphere. 
This formulation provides an intuitive picture of property incompatibility: it vanishes identically for fully aligned or antipodal constellations and reaches its pairwise maximum when the stars are aligned in orthogonal directions.

We established the ($N$-decreasing) upper bound $(E_N)_{\max} = \frac{2}{3}\frac{N}{N-1}$ for $N \ge 3$. 
Crucially, for $N=4$, our analysis reveals a geometric degeneracy where $SU(2)$-inequivalent stellar configurations—such as the regular tetrahedron, a square on a constant-latitude plane, and a non-regular tetrahedron— correspond to the same maximum value $E_4 = 8/9$. 

For larger $N$, an increasing variety of inequivalent maximizing constellations is expected, since the isotropy condition $M = \frac{N}{3}\mathbb{I}_3$ can be satisfied by progressively richer stellar configurations.

To overcome the inapplicability of the Majorana decomposition for higher-dimensional bosons (qudits instead of qubits), we introduced the generalized quantifier $\mathbb{E}_N(\Phi)$. 
It brings together a contribution of the distance of the state of interest to the manifold of symmetrized states via the maximal fidelity, and a contribution from the minimal property-attribution incompatibility measure of the closest symmetrized state.  

Our approach bridges a conceptual gap between the GMW property-based separability criteria and quantum state geometry. 
While the Majorana constellations provide a direct geometrical picture for qubits, the fundamental dependence of the generalized quantifier $\mathbb E_N$ on the pairwise Fubini-Study angles ensures that the geometric formulation of property incompatibility generalizes seamlessly to systems of $d$-level bosons.

Several questions remain open for future investigation. 
For larger numbers of two-level bosons, the characterization of the stellar constellations satisfying the maximal-incompatibility condition may reveal a richer structure of geometrically inequivalent configurations, yet the maximal amount of incompatibility ($E_N)_{\max}$ will decrease progressively as the system's size increases. 
For higher-dimensional systems, it would also be relevant to further characterize the generalized quantifier $\mathbb{E}_N(\Phi)$, in particular the structure of the closest symmetrized states involved in its definition. 
Finally, a fundamental and interesting question is whether the incompatibility of property attribution can be connected to operationally accessible quantum resources in systems of indistinguishable particles. This goal would be relevant for quantifying resources in quantum technologies based on indistinguishable photons, such as photonic quantum computers.

\begin{acknowledgments}
P. C. and A. P. M. acknowledge financial support from Grant No. PIP 11220210100963CO from CONICET (Argentina) and SeCyT, Universidad Nacional de Córdoba. A. V. H. acknowledges financial support by DGAPA-UNAM through project PAPIIT IN110526.  
\end{acknowledgments}

\bibliographystyle{ieeetr} 
\bibliography{bib_entanglement_IP}

\end{document}